\documentclass[twocolumn,english,eng]{revtex4-1}
\usepackage{lmodern}
\usepackage[T1]{fontenc}
\usepackage[latin9]{inputenc}
\usepackage{amstext}
\usepackage{graphicx}
\usepackage{esint}
\usepackage{babel}
\begin{document}

\title{Deformation and chaining of flexible shells in a nematic solvent}

\author{Andrew DeBenedictis}

\affiliation{Department of Physics and Astronomy, Tufts University, 574 Boston
Avenue, Medford, Massachusetts 02155, USA}

\author{Andrea L. Rodarte}

\affiliation{Department of Physics, University of California, Merced, 5200 Lake
Road, Merced, California 95343, USA}

\author{Linda S. Hirst}

\affiliation{Department of Physics, University of California, Merced, 5200 Lake
Road, Merced, California 95343, USA}

\author{Timothy J. Atherton}
\email{timothy.atherton@tufts.edu}

\affiliation{Department of Physics and Astronomy, Tufts University, 574 Boston
Avenue, Medford, Massachusetts 02155, USA}
\begin{abstract}
A micrometer-scale elastic shell immersed in a nematic liquid crystal
may be deformed by the host if the cost of deformation is comparable
to the cost of elastic deformation of the nematic. Moreover, such
inclusions interact and form chains due to quadrupolar distortions
induced in the host. A continuum theory model using finite elements
is developed for this system, using mesh regularization and dynamic
refinement to ensure quality of the numerical representation even
for large deformations. From this model, we determine the influence
of the shell elasticity, nematic elasticity and anchoring condition
on the shape of the shell and hence extract parameter values from
an experimental realization. Extending the model to multi-body interactions,
we predict the alignment angle of the chain with respect to the host
nematic as a function of aspect ratio, which is found to be in excellent
agreement with experiments and greatly improves upon previous theoretical
predictions. 
\end{abstract}
\maketitle

\section{Introduction}

An important application of nematic liquid crystals (LCs) is as guides
for the self assembly of included colloidal particles \cite{meeker2000,musevic2006,lin2008,Koenig2010}.
Chemical treatment of the particles may induce a locally preferred
orientation of the adjacent nematic on their surface and induce elastic
distortions in the bulk liquid crystal. Elasticity-mediated interactions
between particles cause micron-sized particles to self-organize into
chains or clusters that are strongly ($\sim1000k_{B}T$) bound together
\cite{poulin1997,poulin1998,musevic2006,skarabot2008}. Much smaller
nanoparticles (NPs) disperse uniformly in the isotropic phase but
can be sculpted into a variety of structures including networks and
shells by kinetic effects as the host undergoes a transition into
a liquid crystalline phase \cite{meeker2000,milette2012,rodarte2013}.
Self-assembly of nanoparticles in LCs can therefore exploit nucleation
and growth as would occur in an isotropic fluid \cite{miller2014,wu2014},
but the liquid crystalline order permits additional control over the
self-assembled structure \cite{rodarte2013}. 

\begin{figure}
\includegraphics[width=3.3in]{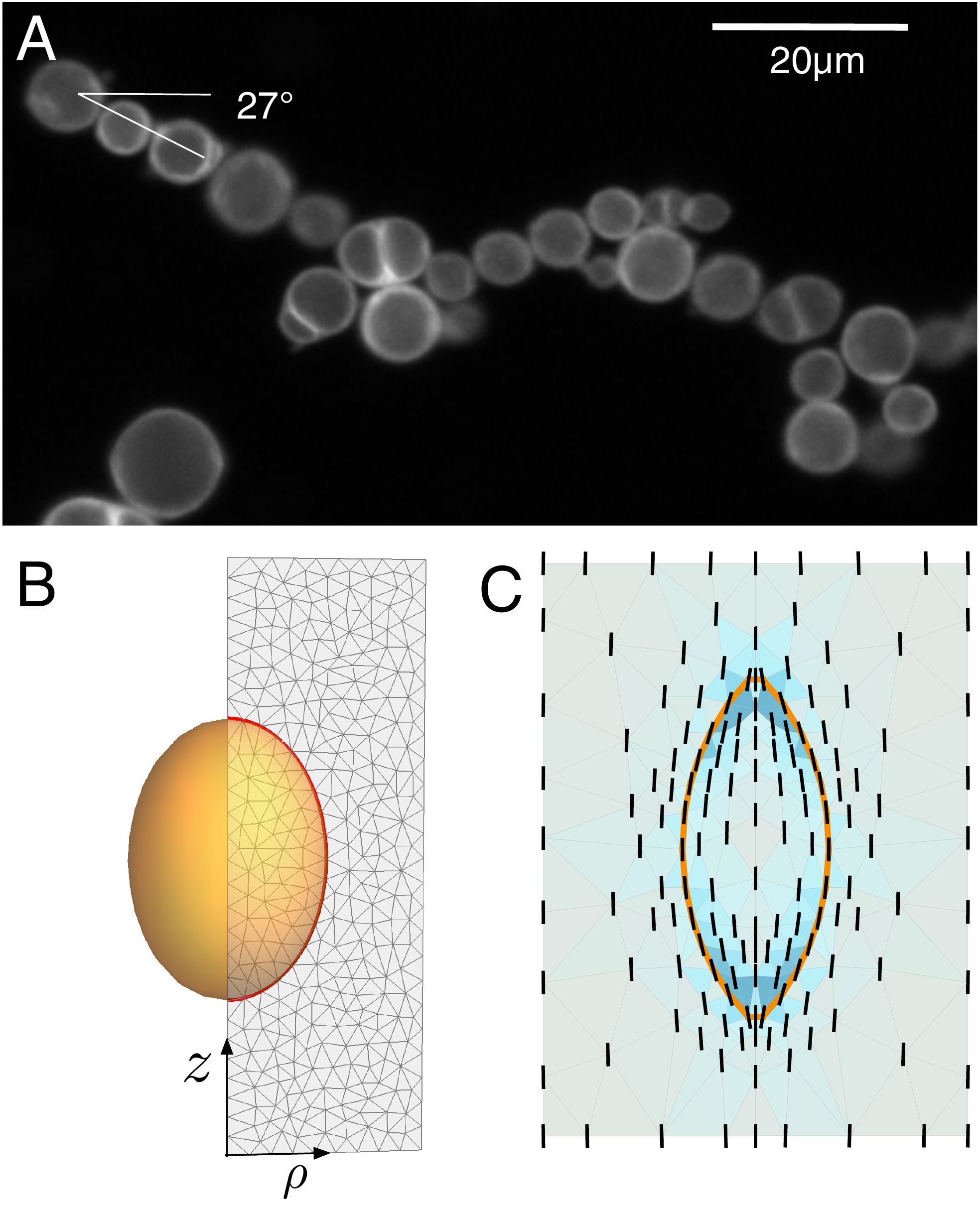}\caption{\label{fig:geom}(Color online)\textbf{ A} Fluorescence microscopy
images of quantum dot microshells\cite{rodarte2015} show subtly elongated
shells that form chains, here at $27^{\circ}$ to the alignment axis
of the nematic host. \textbf{B} Schematic of the computational domain.
\textbf{C} Fully relaxed simulation with $\Gamma_{\sigma}=10$, $\Gamma_{\kappa}=0$,
and $\Gamma_{W}=10$, shaded to indicate the relative energy of each
mesh face. Two copies of the he computational domain are depicted,
one reflected about the $z$ axis, to assist visualization. }
\end{figure}

Hierarchical structures can be formed by combining these mechanisms.
Two co-authors of this paper (Rodarte, Hirst) created nanoparticle
shells \cite{rodarte2015} by cooling a solution of mesogen-functionalized
quantum dots in 5CB from the isotropic to nematic phase. The nanoparticles
are driven to the boundary of the vanishing isotropic phase and by
a nucleating inner nematic domain; they aggregate and solidify, leaving
behind a shell. As shown in Fig. \ref{fig:geom} shells then migrate
after their formation to align in long chains due to the elastic interactions
and, remarkably, adopt a more elongated morphology over time. 

In this paper, we develop a model of the elongation and chaining process
by minimizing the elastic free energy with both with respect to the
spatially varying orientation of the liquid crystal \emph{and} the
shape of the shell. We predict the shape as a function of the elastic
constants for both the nematic and the shell, and determine the orientation
of the chains with respect to the bulk nematic. Results are compared
with experiment and earlier models that neglect the shape of the particles
\cite{poulin1998,Stark2001}. 

Shape-order optimization problems such as the shells in this paper
are challenging because few analytical results are available and computational
approaches must maintain the quality of the numerical representation
during the optimization. A related and widely studied problem of similar
character is is to determine the shape of a \emph{tactoid}, a droplet
of nematic liquid crystal in a host solvent. While a rich variety
of tactoid shapes is observed in lyotropic systems \cite{Zocher1925,Zocher1929,Bernal1941},
including some that are not simply-connected \cite{Casagrande1987,Kim2013},
thermotropic liquid crystals such as 5CB generally form spherical
droplets because of the surface tension between the LC and host tends
to be much larger than the cost of elastic deformations. For the shells
in \cite{rodarte2015} and considered here, significant deformation
occurs because the surface in question is not the interface between
the LC and a host fluid but rather the surface of the solid shell
that is surrounded inside and out by the LC. 

Because no theoretical technique can simultaneously resolve the molecular
scale order of the liquid crystal and the micron-sized shape of the
tactoid, theoretical work requires a trade off between size and resolution.
Prior work includes Monte Carlo methods predict elongation of thermotropic
LCs \cite{Bates2003}, Molecular Dynamics simulations that show elongation
and spontaneous chiral order in nanoscale droplets \cite{mesfin1999,berardi2007,Rull2012,Vanzo2012}
and continuum theory \cite{Kaznacheev2002,Prinsen2004a} that nonetheless
assumes a rigid idealized shape for the tactoid boundary.

Recently \cite{DeBenedictis2016}, we created a finite element continuum
theory model to determine tactoid shape incorporating a new dynamic
mesh control algorithm that ensures the numerical scheme remains accurate
and stable during shape minimization. Here we will adapt the same
strategy to model the shells discussed above. The paper is structured
as follows: in Section \ref{sec:model} the model is described. In
section \ref{sec:shapeSpace} the results are compared to experimental
observations. The chaining phenomenon is studied in section \ref{sec:chains}
using an extension of the model described in section \ref{sec:model}.
Finally, conclusions and future work are presented in Section \ref{sec:conclusion}. 

\section{Model\label{sec:model}}

The system comprises a nanoparticle shell with nematic liquid crystal
on both the interior and exterior. The total free energy comprises
three contributions,
\begin{equation}
F=F_{s}+F_{n}+F_{a},\label{eq:totalenergy}
\end{equation}
i.e. the elastic energy of deforming the shell, the elastic energy
of the nematic and an anchoring term that couples the nematic to the
shell. The shells are composed of nanoparticles stabilized by ligand-ligand
interactions with a very short range attractive attraction\cite{rodarte2015}.
These interactions therefore resist changes in the area of the shell.
Since the shells are only a few nanoparticles thick, we \emph{a priori}
expect the bending energy to be negligible, but nonetheless include
it in the energy to determine its effect on the shape. The shell elastic
contribution to (\ref{eq:totalenergy}) is therefore,
\[
F_{s}=\sigma\int_{\partial S}dA+\kappa\int_{\partial S}\left(H-H_{0}\right)^{2}dA
\]
where $\sigma$ is the surface tension, $H$ is the mean curvature
and $H_{0}$ is a prescribed mean curvature. Because the shells first
form at the interface of an approximately spherical isotropic region
at a critical radius $R$, we assume the preferred mean curvature
$H_{0}$ is that of the initial sphere $\left(1/R\right)$. In practice,
we find that $H_{0}$ and $\kappa$ have covariant effects on the
final shape, so this choice is somewhat arbitrary. 

The elastic energy of the LC is the Frank energy\cite{frank1958},
\begin{equation}
F_{n}=\int_{S,S'}dV\left[K_{1}\left(\nabla\cdot\textbf{n}\right)^{2}+K_{2}\left(\textbf{n}\cdot\nabla\times\textbf{n}\right)^{2}+K_{3}\left|\textbf{n}\times\nabla\times\textbf{n}\right|^{2}\right],\label{eq:frank}
\end{equation}
where $K_{1}$, $K_{2}$, and $K_{3}$ are the splay, twist, and bend
elastic constants and the integral is taken over the entire simulation
volume, i.e. both the interior $S$ and exterior $S'$ of the shell.
A local constraint $\mathbf{n}\cdot\mathbf{n}=1$ is enforced to ensure
the director remains a unit vector. 

Finally, the anchoring term, 
\begin{equation}
F_{a}=W\int_{\partial S}||\textbf{n}-\textbf{n}_{e}||^{2}dA,\label{eq:anchoring}
\end{equation}
imposes a preferred orientation $\mathbf{n}_{e}$ relative to the
shell normal with associated energy $W$. Because the shells are ligand-stabilized,
we additionally impose a volume constraint,
\begin{equation}
\int_{S}dV=V_{0}.\label{eq:vol}
\end{equation}

We non-dimensionalize the problem in the usual way using $\Lambda$,
a length scale of the order of the size of the shell, by changing
variables $x\rightarrow\Lambda x'$ and dividing through by $K_{1}\Lambda$.
Hence, the energy (\ref{eq:totalenergy}) becomes,

\begin{eqnarray}
\frac{F}{K_{1}\Lambda} & = & \Gamma_{\sigma}\int_{\partial S}dA+\Gamma_{\kappa}\int_{\partial S}\left(H'-H'_{0}\right)^{2}dA\nonumber \\
 &  & +F_{n}/K_{1}+\Gamma_{W}\int_{\partial S}||\textbf{n}-\textbf{n}_{e}||^{2}dA.\label{eq:energy-nonD}
\end{eqnarray}
Hence we introduce dimensionless parameters $\Gamma_{\sigma}=\frac{\sigma\Lambda}{K_{1}}$,
$\Gamma_{\kappa}=\frac{\kappa}{K_{1}\Lambda^{3}}$, and $\Gamma_{W}=\frac{W\Lambda}{K_{1}}$,
that represent the relative strengths of the surface tension, mean
squared curvature, and anchoring energy relative to the elastic energy.

The functional (\ref{eq:energy-nonD}) is discretized as follows.
First, we exploit the apparent cylindrical symmetry of the shells
to work in cylindrical polar coordinates $(\rho,\phi,z)$. The computational
domain, shown in Fig. \ref{fig:geom}B, is the $(\rho,z)$ plane that
must be swept out in $\phi$ to recover the full 3D solution and is
discretized into triangular elements. The initially spherical shell
surface is specified as a sequence of edges terminating at top and
bottom on the $\rho=0$ line. The mean curvature $H$ of the shell
at a given vertex on the shell is calculated using the discrete method
from \cite{Dyn:2001:OTU:570828.570841}. Director values are stored
on each vertex and parametrized in cylindrical coordinates, $\textbf{n}=\left(n_{\rho},n_{\phi},n_{z}\right)$
with appropriate derivatives in (\ref{eq:frank}) re-expressed in
these coordinates. Interpolation of the director between vertices
is performed using a special spherical weighted average \cite{buss2001}
that maintains unit length at all points. The Frank energy of each
element is then computed by gaussian quadrature \cite{Dunavant1985}.
The anchoring energy (\ref{eq:anchoring}) is also computed along
the shell by gaussian quadrature. 

Having constructed a finite element approximation to (\ref{eq:energy-nonD}),
we minimize it using gradient descent with respect to both the director
values and vertex positions from an initial state with $\textbf{n}=n_{z}$
at all vertices. To maintain a well-behaved mesh, we supplement the
target functional with auxiliary functionals as described in \cite{DeBenedictis2016}
that promote equiangular elements and uniform energy density between
adjacent elements. Additionally, local refinement and coarsening is
performed to capture adequate detain in regions of high energy density.
Fig. \ref{fig:geom}C shows a converged solution in which the local
refinement and energy density reveal the bipolar field adopted by
the nematic. The system is considered to have converged when the timestep-normalized
percent change to the energy is less than $10^{-6}$ over two cycles
of steps to relax both the vertex location and director orientation.
We explicitly test all solutions for stability by computing the bordered
Hessian matrix $G$ and testing that the number of constraints plus
the number of degrees of freedom for the system is larger than the
negative index of inertia plus the corank of $G$. 

\section{Results}

\subsection{Shell shape\label{sec:shapeSpace}}

\begin{figure}
\includegraphics[width=1\columnwidth]{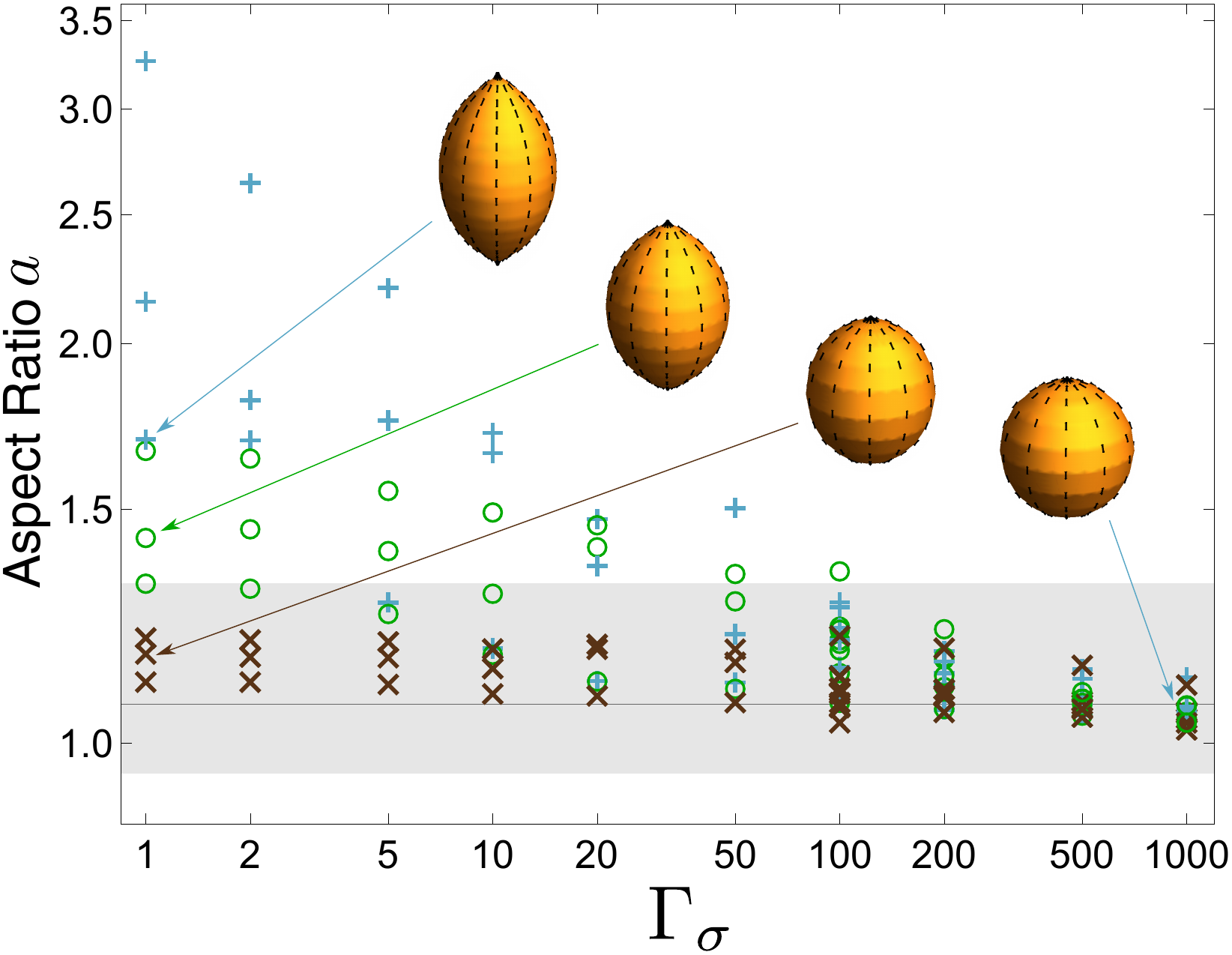}\caption{\label{fig:aspect}Aspect ratio for simulated shells as a function
of surface tension. Cyan crosses, green circles, and brown exes denote
respectively simulations with $\Gamma_{\kappa}=0$, $\Gamma_{\kappa}=0.1$,
and $\Gamma_{\kappa}=1.0$. Variation of aspect ratio within a set
of points with the same $\Gamma_{\sigma}$ and $\Gamma_{\kappa}$
is caused by variation in $\Gamma_{W}$, typically with higher aspect
ratios corresponding to stronger anchoring. Insets visualize the shells
for selected points, all of which have $\Gamma_{W}=100$. The gray
region denotes the range of aspect ratios observed in the experimental
data, with the darker line indicating the median of 1.07.}
\end{figure}

We ran a series of simulations varying the coefficients $\Gamma_{\sigma}$,
$\Gamma_{\kappa}$ and $\Gamma_{W}$ to determine how the final shape
of the shell depends on these parameters and hence identify the space
of shell shapes accessible by this mechanism. One measure of the final
shape is the aspect ratio $a$, which is displayed in Fig. \ref{fig:aspect}
as a function of the surface tension. The high aspect ratios obtained
indicate that the mesh regularization procedure described is effective
in permitting large physical deviations from the initial shape. The
simulations reproduce aspect ratios similar to those observed experimentally
($0.9\leq a\leq1.3$) under two distinct conditions: first, systems
where the mean squared curvature term dominates; second, systems dominated
by surface tension. Either of these cases suggests, as we expect,
that the interactions between ligands of the nanoparticles are considerably
stronger than interactions between LC molecules. Indeed, the experimental
shells remain stable even when the surrounding nematic phase is heated
to isotropic above $34^{\circ}\text{C}$ and remain stable up to about
$100^{\circ}\text{C}$. 

Looking in more detail at the predicted shape of the shells displayed
in the insets of fig. \ref{fig:aspect} there are two distinct morphologies:
Surface tension dominated shells form a characteristic cusp at the
poles; conversely mean squared curvature dominated shells instead
favor smooth poles. The fluorescence microscopy image shown in fig.
\ref{fig:geom}A shows evidence of cusps, and so we conclude that
the shell elasticity is dominated by surface tension. This is expected
a classic result from shell elasticity: the ratio of bend energy to
stretching energy is of order $(R/h)^{2}$ where $R$ is the shell
radius and $h$ is the shell thickness \cite{LandauLifshitz}. We
estimate the shell thickness to be 10 to 100 times smaller than the
radius, which suggests that $\Gamma_{\sigma}/\Gamma_{\kappa}$ lies
between 100 to 10,000. 

Fig. \ref{fig:aspect} therefore allows us to predict the shapes that
would result from changing the experimental system. For example, use
of a LC with larger elastic constants would decrease $\Gamma_{\sigma}$,
resulting in more elongated shells. Similarly, decreasing (increasing)
the concentration of nanoparticles in the initial system would result
in smaller (larger) shells \cite{rodarte2015}, which have a smaller
(larger) $\Gamma_{\sigma}$ and thus will elongate more (less).

\begin{figure}
\includegraphics[width=1\columnwidth]{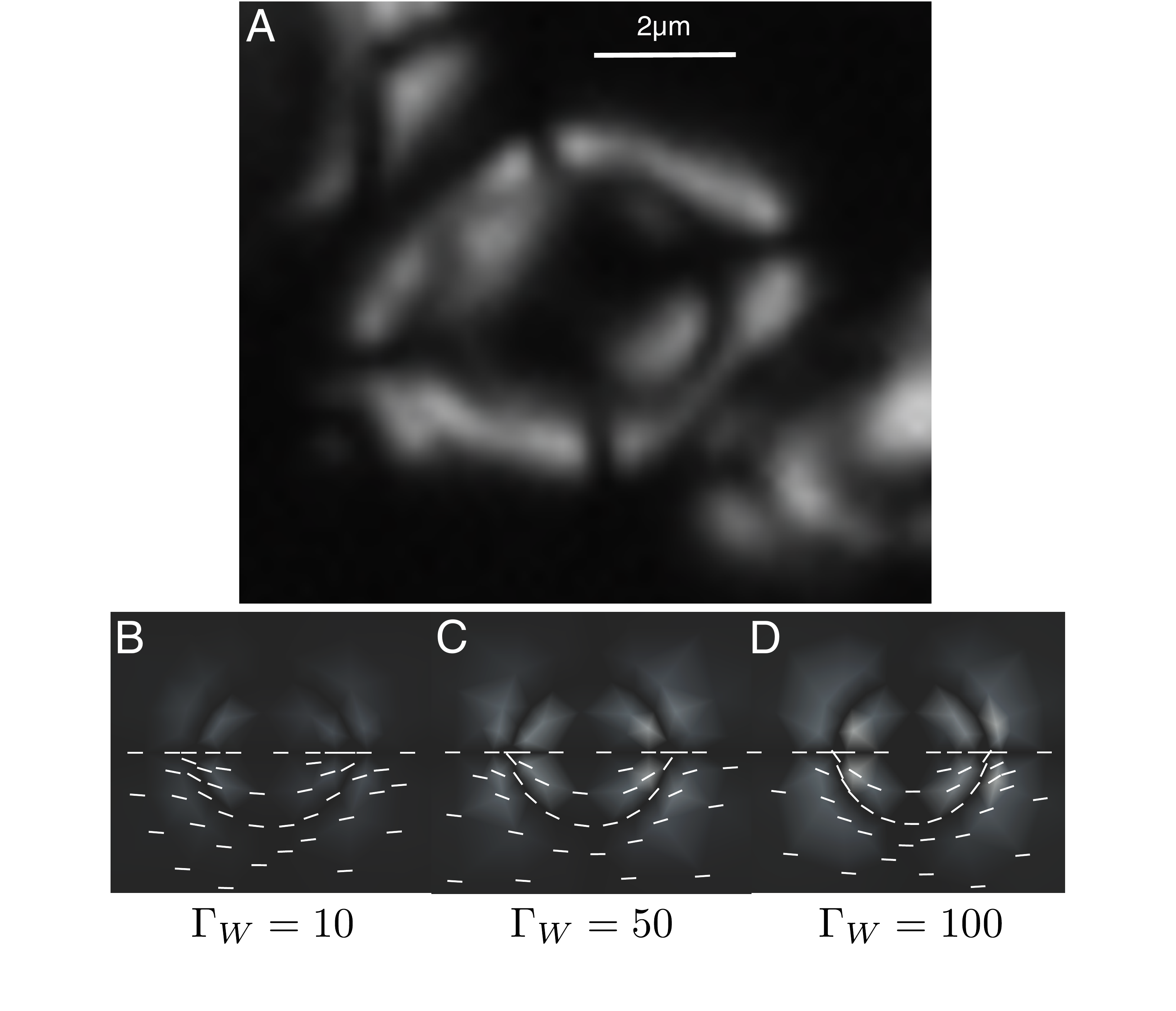}\caption{\label{fig:crossed-polar}\textbf{A} An experimental crossed-polarizer
image of a shell imaged close to the central plane. \textbf{B-D} Simulated
crossed-polarizer images for three shells with different anchoring
strengths ($\Gamma_{W}=10$, $50$, and $100$) and aspect ratios
equal to the median aspect ratio observed experimentally. The director
field superimposed over the simulated images shows stronger anchoring
conditions leading to more curvature in the nematic field.}
\end{figure}

We now turn to the configuration of the liquid crystal around and
inside the shell. To facilitate a comparison with experiment, an approximate
crossed-polarizer microscope image is generated from each configuration.
A very simple optical model is used, treating the LC as a single anisotropic
layer in the $y=0$ plane between two crossed-polarizers above and
below the LC. Thus, the intensity $I$ at a given point is
\begin{equation}
I=\cos^{2}\phi\sin^{2}\phi,\label{eq:polarI-1}
\end{equation}
where $\phi$ is the angle of the director off of the $z$ axis. We
emphasize that this is quite a crude approximation of a true crossed-polarizer
image, using only the 2D slice of the director field. Moreover, the
optical properties of the NP shell itself are unknown, however motivated
by the experimental images, we assume that no light passes through
the pixels that lie directly on the shell boundary at $y=0$.

Results are displayed in panels B through D of Fig. \ref{fig:crossed-polar}
that show simulated microscope images with different anchoring strengths
but aspect ratios all within $2\%$ of the median experimental aspect
ratio. In spite of the very simple optical model, the simulated images
reproduce the main features of the experimental image shown in \ref{fig:crossed-polar}A:
they possess a dark band along the axis of symmetry, indicating LC
alignment along the axis. The dark band perpendicular to this axis
denotes a region of LC alignment parallel to the vertical polarizer.
Bright regions appear inside and outside of the shell where the director
field distorts to match both the anchoring condition at the shell
and either the axis of symmetry or the far-field director orientation.
Weak anchoring requires less perfect matching of the easy axis at
the shell, and thus less LC field deformation and dimmer images. The
similarity between the two strongest anchoring cases and the experimental
image suggests anchoring strength of at least $\Gamma_{W}=50$. This
corresponds to a value of $W$ of order $10^{-5}\text{J/m}^{2}$,
consistent with characterizations of strong anchoring \cite{Kuksenok1996,Ruhwandl1997}.
That cases with strong anchoring require $\Gamma_{\sigma}\ge500$
to achieve aspect ratios of 1.07 is consistent with the surface tension-dominated
system.

\subsection{Shell-shell interactions and chaining\label{sec:chains}}

\begin{figure*}
\includegraphics[width=2.05\columnwidth]{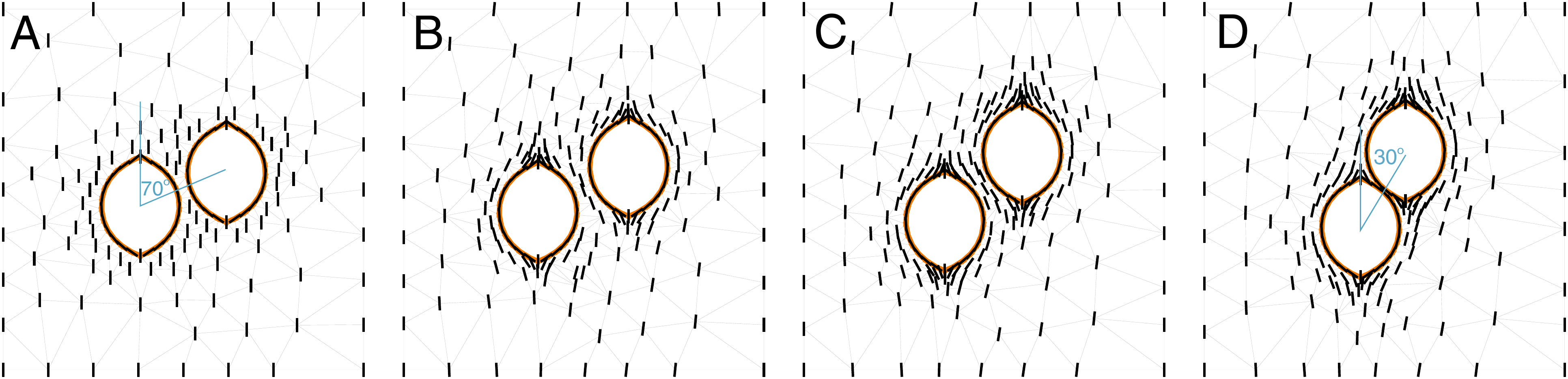}\caption{\label{fig:chains}\textbf{A-D} Simulation snapshots of two shells
with aspect ratio = 1.18 relaxing from $\theta_{0}=70^{\circ}$ to
$\theta_{p}=30^{\circ}$. The shells repel initially, before finding
their preferred alignment angle and attracting one another.}
\end{figure*}

After formation, the NP shells are observed to migrate over the course
of minutes to form chains aligned at some angle to the host nematic
(Fig. \ref{fig:geom}A). Previous studies with colloidal glass spheres
instead of shells behave similarly, as this feature is generic to
particles that produce a quadrupolar distortion field in the host
nematic \cite{musevic2006,Ruhwandl1997,Smalyukh2005}. These authors
report observed alignment angles of $\sim30^{\circ}$ for particles
with parallel anchoring, while our NP shells align at angles between
$25^{\circ}$ and $35^{\circ}$ (Fig. \ref{fig:geom}A). Recently,
chaining has also been observed in ferromagnetic nanoparticles\cite{dierking2017ordering},
although the chaining angles observed there are somewhat higher $\sim54^{\circ}$
which is likely because of the presence of shell-shell interactions
other than elasticity. 

Prior theoretical calculations predict much larger chaining angles,
between $45^{\circ}$ and $49^{\circ}$, than those observed either
with shells or microspheres. This is because they take a far-field
approximation and neglect short-range distortions induced by the particle
shape \cite{Ramaswamy1996,poulin1998,Araki2006,Stark2001}. Here,
we will adapt the shape evolution technique developed above to include
two shells that are mobile and hence able to locate an equilibrium
state. Because the shells composing these chains do not share axes
of symmetry, the cylindrical domain above cannot be used. Instead,
we develop a two-dimensional model that includes only the projections
of the shells onto the $y=0$ plane.

\begin{figure}
\includegraphics[width=1\columnwidth]{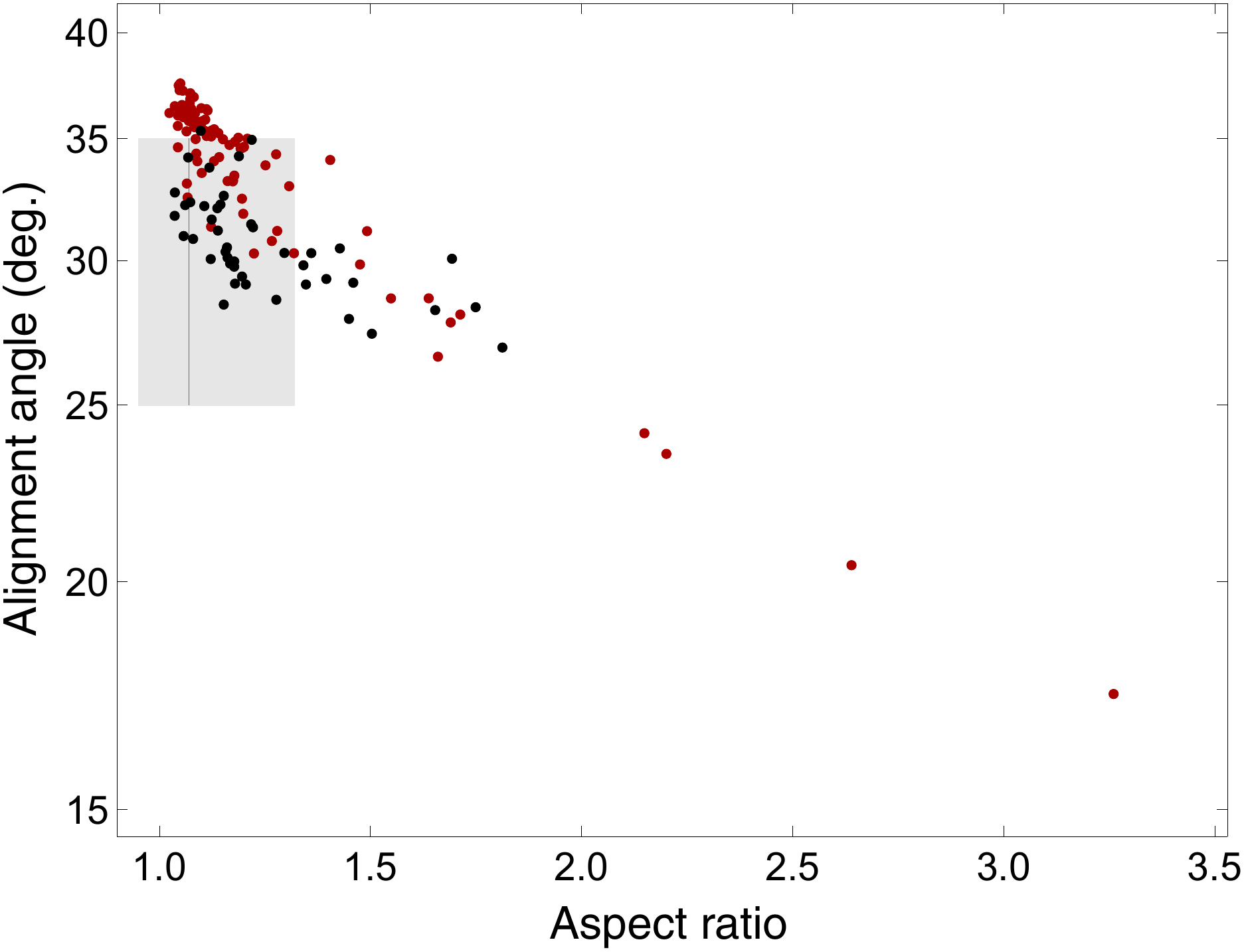}\caption{\label{fig:chainPlot} (Color online) Preferred alignment angle $\theta_{p}$
as a function of shell aspect ratio for all of the simulated cases
shown in fig. \ref{fig:aspect}. The black series contains cases with
strong anchoring ($\Gamma_{W}>50$) while the red series contains
cases with weaker anchoring ($\Gamma_{W}\le50$). The gray region
denotes the range of aspect ratios and angles observed in the experimental
data, with the darker line indicating the median of 1.07.}
\end{figure}

Minimization of elastic deformations in the host nematic drives the
alignment of the shells. However, from the results in section \ref{sec:shapeSpace}
these elastic forces are weak compared to the surface tension and
anchoring forces that define the shapes of the shells and the LC orientation
at the shell-nematic interfaces. The shape of the shells and anchoring
conditions are therefore fixed and so the only the elasticity term
of (\ref{eq:totalenergy}) is minimized with respect to the shape
of the host nematic domain. 

To initialize the simulation, shell shapes and LC director fields
at the LC-shell interface are taken from final states of the simulations
described in section \ref{sec:shapeSpace} to use in the chaining
model. For each run, a mesh is built containing two such shells with
their axes of symmetry along the $z$ axis, but offset such that their
centers form an angle $\theta_{0}$ with this axis. Again, we take
gradient descent steps alternatingly to relax the vertex locations,
director orientations, and mesh quality. As the anchoring of the LC
to the shell boundary is fixed, the interior regions of the shells
does not contribute to the alignment and is not included in the computational
domain. The preferred alignment angle $\theta_{p}$ does not depend
on $\theta_{0}$. 

Results from a typical run are shown in Fig. \ref{fig:chains}. Panels
A through D show selected snapshots from a simulation where the starting
alignment angle $\theta_{0}=70^{\circ}$. Initially, the shells repel
each other, before rotating around to find a preferable angle. Once
at this angle, the space between the shells collapses as nematic forces
attract them. This behavior of repulsion at angles close to $0^{\circ}$
or $90^{\circ}$ and attraction at intermediate angles is well-documented
experimentally \cite{Ruhwandl1997,Smalyukh2005}. 

Fig. \ref{fig:chainPlot} displays the calculated chaining angle as
a function of the aspect ratio of the NP shells with strong (black)
and weak (red) anchoring. As aspect ratio increases, the chaining
angle is reduced. Noise in the plot is due to variations in the anchoring
condition and bend modulus from the initial configuration. For shells
with aspect ratios in the range observed experimentally, we see preferred
angles mostly from $28^{\circ}$ to $36^{\circ}$, which is in very
good agreement with the chains seen experimentally. Furthermore, extrapolation
of the strong anchoring data in the limit of $a\to1$, i.e. spherical
colloidal particles with rigid anchoring, $\theta_{p}=32$, which
agrees very well with the $30^{\circ}$ reported by other authors
than any previous method.

\section{Conclusion\label{sec:conclusion}}

We present a continuum theory finite element model for deformation
and two body interactions of a flexible shell in a nematic liquid
crystal. The model features dynamic mesh remodeling utilizing auxiliary
functionals to maintain accuracy despite large deformations from the
initial configuration. The model is used to simulate experimentally
observed elongation and chaining of mesogen-functionalized nanoparticle
shells that form as the LC is quenched into the nematic phase. Because
the elastic behavior of these shells is unknown, the model enables
us to extract the relative contribution of shell elasticity, nematic
elasticity and anchoring from the observed shapes. By comparing simulations
with experimental images, we determine surface tension and anchoring
dominate the shell shape.

Extending the model to incorporate multiple shells, we predict the
chaining angles attained by the shells, and determine the dependence
of the chaining angles on the aspect ratios of the shells. Furthermore,
our model predicts the chaining angle of spherical particles with
strong planar anchoring in a nematic LC far more accurately than any
previous theoretical treatment.

Our simulations cover a wide range of parameter space, so information
from these results enables us to design systems with particular shell
sizes, shapes, and alignment angles. This ability to control and tune
particle shape by an highly scalable self-assembly method is valuable
as part of designing hierarchical processes. Furthermore, the simulation
methodology presented is computationally cheap and readily adapted,
with little modification, to a wide variety of shape-order problems
involving soft materials. 
\begin{acknowledgments}
\emph{ADB is supported by a Tufts University Burlingame Fellowship.
TJA acknowledges support from the National Science Foundation under
grant no. }DMR-1654283\emph{. TJA is also supported by a Cottrell
Award from the Research Corporation for Science Advancement. LSH and
ALR acknowledge financial support from the National Science Foundation
grant no. }DMR-CBET-1507551\emph{. This work was partly performed
at the Aspen Center for Physics, which is supported by National Science
Foundation grant }PHY-1066293\emph{.}
\end{acknowledgments}

\end{document}